\documentclass[english]{article}
\usepackage[T1]{fontenc}
\usepackage[latin1]{inputenc}
\usepackage{geometry}
\usepackage{setspace}
\makeatletter

 \newcommand{\lyxaddress}[1]{
   \par {\raggedright #1 
   \vspace{1.4em}
   \noindent\par}
 }

\usepackage{setspace}

\usepackage{graphicx}
\usepackage{achemso}
\usepackage{overcite}
\usepackage{babel}

\usepackage{babel}
\makeatother
\begin{document}

\title{Diverse Spreading Behavior of Binary Polymer Nanodroplets}

\author{David R. Heine, Gary S. Grest, Edmund B. Webb III}

\maketitle

\lyxaddress{\begin{center}Sandia National Laboratories, Albuquerque, New Mexico
87185\end{center}}

\begin{abstract}
Molecular dynamics simulations are used to study the spreading of
binary polymer nanodroplets in a cylindrical geometry. The polymers,
described by the bead-spring model, spread on a flat surface with
a surface-coupled Langevin thermostat to mimic the effects of a corrugated
surface. Each droplet consists of chains of length 10 or 100 monomers
with $\sim350\,000$ monomers total. The qualitative features of the
spreading dynamics are presented for differences in chain length,
surface interaction strength, and composition. When the components
of the droplet differ only in surface interaction strength, the more
strongly wetting component forms a monolayer film on the surface even
when both materials are above or below the wetting transition. In
the case where the only difference is the polymer chain length, the
monolayer film beneath the droplet is composed of an equal amount
of short chain and long chain monomers even when one component (the
shorter chain length) is above the wetting transition and the other
is not. The fraction of short and long chains in the precursor foot
depends on whether or not both the short and long chains are in the
wetting regime. Diluting the concentration of the strongly wetting
component in a mixture with a weakly wetting component decreases the
rate of diffusion of the wetting material to the surface and limits
the spreading rate of the precursor foot, but the bulk spreading rate
is not affected until the more strongly wetting component is removed
completely. 
\end{abstract}

\section{Introduction}

The spreading of liquid droplets on a surface is an important issue
for several industries including adhesion, lubrication, coatings,
and printing. Emerging nanotechnology in areas such as lithography
and microfluidics make the issue of droplet spreading on small length
scales even more relevant.

Just as blending bulk polymers can improve the physical properties
of the resulting material, adding a second component to a spreading
droplet can produce a desired change in surface tension or wettability.
For example, adding a surfactant to a droplet that ordinarily does
not wet a surface can give a product that does wet the surface.\cite{KCB:CS:90}
In general, adding a second component provides more parameters with
which to tune the material properties, but it also introduces more
complex phenomena such as the interdiffusion of the two components
and possible preferential wetting of one at either the liquid/solid
or liquid/vapor interface due to the difference in surface tension.

Compared to homogeneous systems, the published literature on binary
droplets is limited. Experiments on binary droplets have been concerned
primarily with a nonwetting polymer solute in a wetting solvent. These
experiments explored the concentration dependence of the equilibrium
contact angle\cite{RAR:MAC:92} and found a leak-out transition\cite{FB:EL:97,BFB:IJE:00}
where a film of pure solvent is in equilibrium with the droplet. Other
experimental\cite{SKE:SCI:92} and theoretical papers\cite{PIR:JSP:03,YC:JCP:03}
focus on the equilibrium behavior of binary mixtures. The surface
segregation of two-component polymer films has been studied using
molecular dynamics (MD) simulation\cite{LBC:JCP:98,T:PRL:99,ABS:POL:04}
as well as numerical techniques based on Ising-like models.\cite{PB:PRA:92,KKB:PRL:96,PB:PRL:01}
Some MD simulations of droplet spreading have been performed for mixtures
of solvent and oligomers\cite{NA:EL:94,NA:PRE:94} with $4\,000$
monomers or mixtures of two oligomers\cite{VRC:Lan:00} with up to
$25\,000$ monomers, though even larger system sizes are required
to adequately model the spreading dynamics. One work by the current
authors explored reactive wetting of binary droplets in metallic systems.\cite{WHG:AM:05}
Results therein demonstrated a difference in substrate dissolution
rate with varying droplet composition, which in turn resulted in different
wetting kinetics and equilibria. However, the combined phenomena of
substrate dissolution with wetting prevents directly connecting those
results to non-reactive wetting of binary polymer droplets.

This paper presents molecular dynamics simulations of binary polymer
nanodroplets differing in either polymer chain length or surface interaction
strength. We explore the qualitative differences in the spreading
behavior, surface monomer composition, and precursor foot composition
for several systems. We also vary the relative concentration of the
two components and analyze the velocity distributions of the spreading
droplets. All of the simulations presented here are for droplets of
$\sim350\,000$ monomers, which for single component droplets is sufficiently
large to overcome finite size effects.\cite{HGW:PRE:03,HGW:PRE:04}

This paper is organized as follows. Section \ref{sec:details} presents
a brief summary of the molecular dynamics simulation techniques and
the application of the surface-coupled Langevin thermostat. It also
describes the methods used to analyze the simulation results. The
methods of analysis are described in more detail elsewhere.\cite{HGW:PRE:03,HGW:PRE:04}
Section \ref{sec:results} contains the simulation results for binary
polymer nanodroplets for different polymer chain lengths and surface
interaction strengths, and discusses the effects of varying the mole
fraction of strongly wetting and weakly wetting components. Conclusions
drawn from these simulations are presented in Sec. \ref{sec:conclusions}.

\section{Simulation Details\label{sec:details}}

Molecular dynamics simulations are performed using the standard coarse-grained
model for polymer chains in which the polymer is represented by spherical
beads of mass $m$ attached by springs. We use a truncated Lennard-Jones
(LJ) potential to describe the interaction between the monomers. The
LJ potential is given by

\begin{equation}
U_{LJ}(R)=\left\{ \begin{array}{rl}
4\varepsilon\left[\left(\frac{\sigma}{r}\right)^{12}-\left(\frac{\sigma}{r}\right)^{6}\right] & r\leq r_{c}\\
0 & r>r_{c}\end{array}\right.\label{eq:ljcut}\end{equation}
 where $\varepsilon$ and $\sigma$ are the LJ units of energy and
length and the cutoff is set to $r_{c}=2.5\:\sigma$. The monomer-monomer
interaction $\varepsilon$ is used as the reference and all monomers
have the same diameter $\sigma$. For bonded monomers, we apply an
additional potential where each bond is described by the finite extensible
nonlinear elastic (FENE) potential\cite{KG:JCP:90} with $k=30\:\varepsilon$
and $R_{0}=1.5\:\sigma$.

The substrate is modeled as a flat surface since it was found previously\cite{HGW:PRE:03}
that with the proper choice of thermostat, the simulations using a
flat surface exhibit the same behavior as those using a realistic
atomic substrate. Since simulating a realistic substrate requires
several times the total number of atoms in the simulation, using the
flat surface greatly improves the computational efficiency. The interactions
between the surface and the monomers in the droplet at a distance
$z$ from the surface are modeled using an integrated LJ potential,

\begin{equation}
U_{LJ}^{wall}(z)=\left\{ \begin{array}{rl}
\frac{2\pi\varepsilon_{w}}{3}\left[\frac{2}{15}\left(\frac{\sigma}{z}\right)^{9}-\left(\frac{\sigma}{z}\right)^{3}\right] & z\leq z_{c}\\
0 & z>z_{c}\end{array}\right.\label{eq:ljwall}\end{equation}
 where $\varepsilon_{w}$ is the monomer-surface interaction strength
and $z_{c}=2.2\sigma.$

We apply the Langevin thermostat to provide a realistic representation
of the transfer of energy in the droplet. The Langevin thermostat
simulates a heat bath by adding Gaussian white noise and friction
terms to the equation of motion,

\begin{equation}
m_{i}\mathbf{\ddot{r}}_{i}=-\Delta U_{i}-m_{i}\gamma_{L}\dot{\mathbf{r}_{i}}+\mathbf{W}_{i}(t),\label{eq:lang}\end{equation}
 where $m_{i}$ is the mass of monomer $i$, $\gamma_{L}$ is the
friction parameter for the Langevin thermostat, $-\Delta U_{i}$ is
the force acting on monomer $i$ due to the potentials defined above,
and $\mathbf{W}_{i}(t)$ is a Gaussian white noise term.\cite{GK:PRA:86}
Coupling all of the monomers to the Langevin thermostat has the unphysical
effect of screening the hydrodynamic interactions in the droplet and
not damping the monomers near the surface stronger than those in the
bulk. To overcome this, we use a Langevin coupling term with a damping
rate that decreases exponentially away from the substrate.\cite{BP:PRE:01}
We choose the form $\gamma_{L}(z)=\gamma_{L}^{s}\exp\left(\sigma-z\right)$
where $\gamma_{L}^{s}$ is the surface Langevin coupling and $z$
is the distance from the substrate. We generally use values of $\gamma_{L}^{s}=10.0\,\tau^{-1}$
and $3.0\,\tau^{-1}$ for $\varepsilon_{w}=2.0\,\varepsilon$ and
$3.0\,\varepsilon$, respectively, based on earlier work\cite{HGW:PRE:03}
matching the diffusion constant of the precursor foot for flat and
atomistic substrates. The larger $\gamma_{L}^{s}$ corresponds to
an atomistic substrate with larger corrugation and hence larger dissipation
and slower diffusion near the substrate.

All of the droplets presented here are modeled as hemicylinders as
described previously.\cite{HGW:PRE:04} The droplets spread in the
$x$ direction and each system is periodic in the $y$ direction with
length $L_{y}=40\,\sigma$ and open in the other two directions. This
allows a larger droplet radius to be studied in the cylindrical geometry
than in the spherical geometry using the same number of monomers.
The binary droplets consist of either mixtures of chain length $N=10$
and $N=100$ polymers with the same surface interaction strength $\varepsilon_{w}$
or mixtures of chain length $N=10$ with varying $\varepsilon_{w}$.
Unless explicitly stated, the binary droplets contain an equal number
of monomers of each component with initial droplet radius of $R_{0}=80\,\sigma$
for a total size of $\sim350\,000$ monomers. These droplets are large
enough that finite droplet size effects are minimal.\cite{HGW:PRE:03}

The equations of motion are integrated using a velocity-Verlet algorithm.
We use a time step of $\Delta t=0.01\;\tau$ where $\tau=\sigma\left(\frac{m}{\varepsilon}\right)^{1/2}$.
The simulations are performed at a temperature $T=\varepsilon/k_{B}$
using the \textsc{lammps} code.\cite{P:JCP:95} Most of the simulations
were run on $48$ to $64$ processors of Sandia's ICC Intel Xeon cluster.
One million steps for a wetting drop of $350\,000$ monomers takes
$37$ hours on $48$ processors.

For the simulations presented here, the instantaneous contact radii
for the precursor foot and bulk regions are extracted every $400\,\tau$.
The contact radius is calculated by defining a one-dimensional radial
distribution function, as described previously.\cite{HGW:PRE:03}
The precursor foot radius calculation includes all monomers that are
within $1.5\,\sigma$ of the surface whereas the bulk radius calculation
uses monomers between $4.5$ and $6.0\,\sigma$ from the surface.

\section{Results and Discussion\label{sec:results}}

The additional degrees of freedom for binary droplets leads to a number
of qualitatively distinct spreading characteristics. In the following
figures, we show snapshots of several different droplets as they wet
the substrate. In each case, the droplets start in roughly the same
configuration with a contact angle near $90^{o}$, but they each show
differences in the precursor foot composition and spreading rate as
well as the composition of the first layer above the substrate.

\subsection{Substrate Interaction Strength}

\begin{figure}
\begin{center}\includegraphics[%
  clip,
  width=7in,
  keepaspectratio]{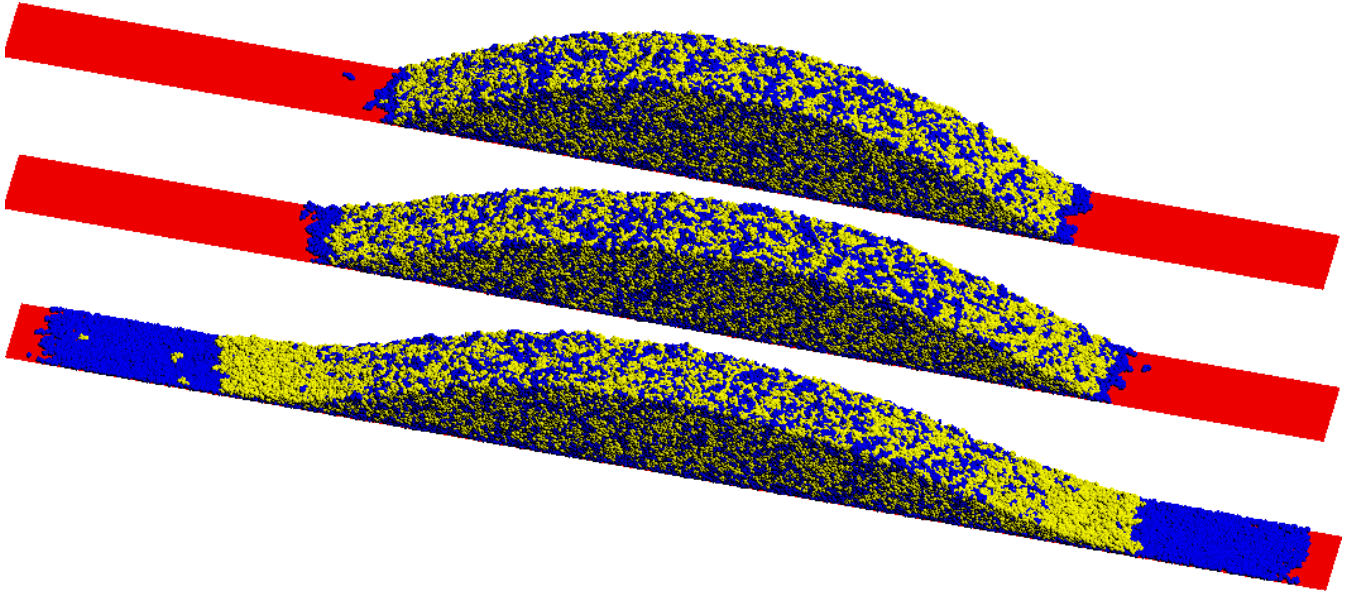}\end{center}

\caption{\label{cap:conf12} Profiles for binary droplets composed of equal
mixtures of chain length $N=10$ with $\varepsilon_{w}=1.0\,\varepsilon$
(yellow) and $\varepsilon_{w}=2.0\,\varepsilon$ (blue) at $t=40\,000\,\tau$
(top) and $t=80\,000\,\tau$ (middle). Profile for $\varepsilon_{w}=2.0\,\varepsilon$
(yellow) and $\varepsilon_{w}=3.0\,\varepsilon$ (blue) at $t=40\,000\,\tau$
(bottom). The substrate (red) has a length of $600\,\sigma$ and a
width of $40\,\sigma$ in each profile. $\gamma_{L}^{s}=10.0\,\tau^{-1}$.}
\end{figure}

\begin{figure}
\begin{center}\includegraphics[%
  clip,
  width=3.25in,
  keepaspectratio]{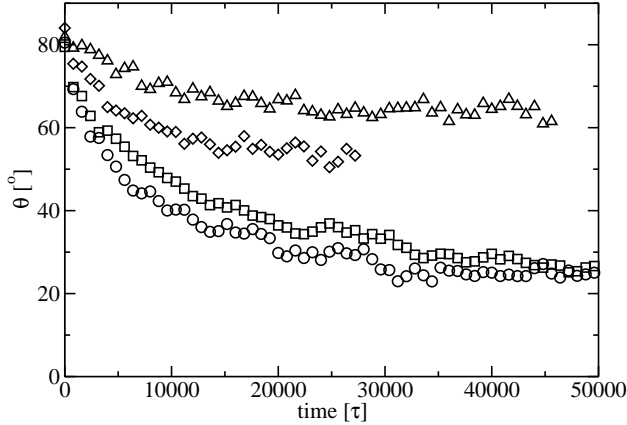}\end{center}

\caption{\label{cap:binangle} Dynamic contact angle for the binary systems
with an equal mixture of $\varepsilon_{w}=0.5\,\varepsilon$ and $\varepsilon_{w}=1.5\,\varepsilon$
(triangles) and an equal mixture of $\varepsilon_{w}=1.0\,\varepsilon$
and $\varepsilon_{w}=2.0\,\varepsilon$ (squares) compared to the
single component droplets with $\varepsilon_{w}=1.5\,\varepsilon$
(diamonds) and $\varepsilon_{w}=2.0\,\varepsilon$ (circles). $N=10$,
$\gamma_{L}^{s}=10.0\,\tau^{-1}$.}
\end{figure}

The top two droplets in Fig. \ref{cap:conf12} show a profile of a
spreading droplet for the mixture of chain length $N=10$ in which
half of the chains interact with the substrate with $\varepsilon_{w}=1.0\,\varepsilon$
and half with $\varepsilon_{w}=2.0\,\varepsilon$ for two different
times. Previously we have shown that the wetting transition for droplets
of chain length $N=10$ occurs at $\varepsilon_{w}^{c}\simeq1.75\,\varepsilon$.
This droplet is a mixture of a wetting and nonwetting polymer.\cite{HGW:PRE:03}
For $N=10$, $\varepsilon_{w}=1.0\,\varepsilon$ has a finite contact
angle of $\theta_{0}\cong90^{o}$. At early times, a monolayer of
the wetting component forms at the solid interface and wets the substrate.
However, unlike the case of a homogeneous droplet in which all monomers
interact with the substrate with $\varepsilon_{w}=2.0\,\varepsilon$\cite{HGW:PRE:04},
the precursor foot in this case does not separate from the main droplet.
Both components of the droplet follow the monolayer as it wets the
substrate, but the spreading rate of the droplet is limited by the
spreading rate of the monolayer. The evolution of the contact angle
for the binary droplet with $\varepsilon_{w}=1.0\,\varepsilon$ and
$\varepsilon_{w}=2.0\,\varepsilon$ is compared to the single component
droplet with $\varepsilon_{w}=2.0\,\varepsilon$ in Fig. \ref{cap:binangle}.
In this case, adding the lower $\varepsilon_{w}$ component slightly
decreases the spreading rate of the droplet. This decrease is sufficient
to produce a better fit to the hydrodynamic spreading model\cite{HGW:PRE:04}
as compared to the pure $\varepsilon_{w}=2.0\,\varepsilon$ system.

Different behavior is observed when both components wet the surface
as shown in the bottom panel of Fig. \ref{cap:conf12} for a mixture
with $\varepsilon_{w}=2.0\,\varepsilon$ and $3.0\,\varepsilon$.
Despite the fact that both components wet the substrate, a monolayer
of the $\varepsilon_{w}=3.0\,\varepsilon$ component still forms at
the solid interface since that is the strongly favored interaction.
Although earlier simulations showed an accumulation of the more strongly
wetting component in the first layer above the substrate,\cite{VRC:Lan:00}
here the first layer consists entirely of the more strongly wetting
component. Unlike the $\varepsilon_{w}=1.0\,\varepsilon/2.0\,\varepsilon$
case, the spreading rate of the precursor foot is fast enough to allow
it to advance well ahead of the bulk of the droplet similar to the
leak-out behavior observed experimentally for polymer/solvent mixtures.\cite{FB:EL:97,BFB:IJE:00}
A depletion region forms near the edges of the bulk droplet because
the diffusion rate of the wetting component in the bulk is insufficient
to replace the material forming the precursor foot. In Section \ref{sub:Varying-Composition}
we return to this case for varying composition of the two components.

We also consider the case of a droplet containing a mixture of two
nonwetting components, $\varepsilon_{w}=0.5\,\varepsilon$ and $\varepsilon_{w}=1.5\,\varepsilon$
for $N=10$. As in the previous cases, a monolayer of the more wetting
component forms between the droplet and the substrate. In this case,
the droplet reaches an equilibrium contact angle of $\theta_{0}\cong64^{o}$
after $20\,000\,\tau$. The evolution of the contact angle for this
nonwetting system is compared to the $\varepsilon_{w}=1.5\,\varepsilon$
single component system in Figure \ref{cap:binangle}. The single
component droplets have equilibrium contact angles of $\theta_{0}\cong120^{o}$
for $\varepsilon_{w}=0.5\,\varepsilon$ and $44^{o}$ for $\varepsilon_{w}=1.5\,\varepsilon$.
Thus the contact angle of the binary system is more strongly influenced
by the material that forms a monolayer on the substrate.

\subsection{Polymer Chain Length}

\begin{figure}
\begin{center}\includegraphics[%
  clip,
  width=7in,
  keepaspectratio]{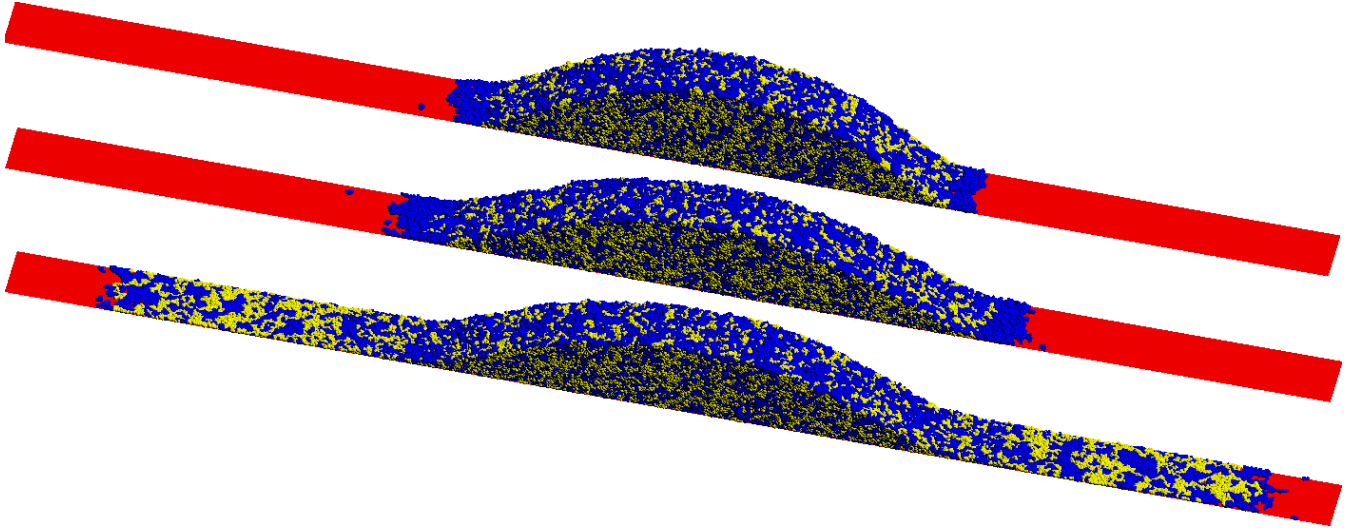}\end{center}

\caption{\label{cap:conf210} Profiles for binary droplets composed of equal
number of monomers of chain length $N=10$ (blue) and $N=100$ (yellow)
polymers with $\varepsilon_{w}=2.0\,\varepsilon$ and $\gamma_{L}^{s}=10.0\,\tau^{-1}$
at $t=40\,000\,\tau$ (top) and $t=80\,000\,\tau$ (middle). Same
as above with $\varepsilon_{w}=3.0\,\varepsilon$ and $\gamma_{L}^{s}=3.0\,\tau^{-1}$
taken at $t=40\,000\,\tau$ (bottom). The substrate (red) has a length
of $800\,\sigma$ and a width of $40\,\sigma$ in each profile.}
\end{figure}

For droplets composed of two chain lengths of the same type of polymer,
the spreading behavior is qualitatively different. The top two droplets
in Fig. \ref{cap:conf210} show profiles for the mixture of chain
lengths $N=10$ and $100$ with $\varepsilon_{w}=2.0\,\varepsilon$
and $\gamma_{L}^{s}=10.0\,\tau^{-1}$. Although $\varepsilon_{w}=2.0\,\varepsilon$
for both components, the longer chains are in the nonwetting regime
($\varepsilon_{w}^{c}\simeq2.25\,\varepsilon$)\cite{HGW:PRE:04}
whereas the shorter chains are slightly above the wetting transition
($\varepsilon_{w}^{c}\simeq1.75\,\varepsilon$). As a result, a slow
precursor foot consisting of approximately 80\% $N=10$ monomers and
20\% $N=100$ monomers extends from the bulk region. This is somewhat
similar to the {}``leak-out'' behavior observed experimentally in
mixtures of polymer and solvent where the precursor foot consists
entirely of solvent.\cite{FB:EL:97,BFB:IJE:00} Unlike Fig. \ref{cap:conf12},
a monolayer of the wetting component does not form on the substrate
below the bulk region of the droplet. Instead, the first monolayer
at the substrate beneath the droplet consists of an equal fraction
of the two components. This can be understood by noting that the longer
chains are nonwetting in this case due to their greater surface tension.
The chains that are buried beneath the droplet feel no influence of
the liquid/vapor interface, so there is no preference as to which
chain length is in contact with the substrate. The chains at the upper
surface of the droplet, both in the bulk region and the precursor
foot, are influenced by the liquid/vapor interface and there one finds
an abundance of the shorter chains. The liquid/vapor surface tensions
for the $N=10$ and $N=100$ chains were previously found to be $0.84$
and $0.96\pm0.02\,\varepsilon/\sigma^{2}$, respectively.\cite{HGW:PRE:04}
The surface tension for the $10/100$ mixture was found to be $0.90\pm0.02\,\varepsilon/\sigma^{2}$
indicating a roughly equal influence by each component on the surface
tension of the mixture. Thus, the surface tension calculation is not
precise enough to account for the abundance of shorter chains at the
liquid/vapor interface.

If $\varepsilon_{w}$ is increased so that both chain lengths are
in the wetting regime, the behavior is similar to that of a single-component
droplet. This is shown in the bottom of Fig. \ref{cap:conf210} for
the mixture of chain lengths $N=10$ and $100$ where $\varepsilon_{w}=3.0\,\varepsilon$
and $\gamma_{L}^{s}=3.0\,\tau^{-1}$. Here, both chain lengths are
in the wetting regime and the rapidly spreading precursor foot is
composed of an equal number of monomers of the two chain lengths.
One consequence of this is the $N=100$ chains diffuse across the
substrate more rapidly when mixed with $N=10$ chains than in the
pure $N=100$ droplet. Although the longer chains have a higher surface
tension, the equilibrium state for both components is a $0^{o}$ contact
angle so there is no segregation of the two components in the precursor
foot as this droplet spreads. The lack of segregation for a mixture
of two chain lengths with the same $\varepsilon_{w}$ was previously
reported by Vou\'{e} \textit{et al.} for chain lengths $8$ and $16$.\cite{VRC:Lan:00} 

\begin{figure}
\begin{center}\includegraphics[%
  clip,
  width=3.25in
  ]{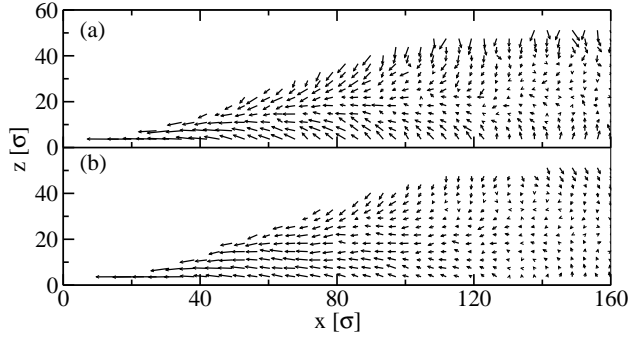}\end{center}

\caption{\label{cap:vel100} Velocity profiles of half of the droplet showing
the $N=10$ chain length (a), and the $N=100$ chain length (b), for
the binary droplet with $\varepsilon_{w}=3.0\,\varepsilon$ and $\gamma_{L}^{s}=3.0\,\tau^{-1}$
at $t=20\,000\,\tau$.}
\end{figure}

The droplet velocity profiles provides another method to analyze the
dynamics of the spreading droplet. Figure \ref{cap:vel100} shows
the velocity profiles for the mixture of chain lengths $10$ and $100$.
Although both components show the same spreading pathway, the shorter
chain lengths tend to move slightly faster at the liquid/vapor and
liquid/solid interfaces.

\subsection{Varying Composition\label{sub:Varying-Composition}}

\begin{figure}
\begin{center}\includegraphics[%
  clip,
  width=7in,
  keepaspectratio]{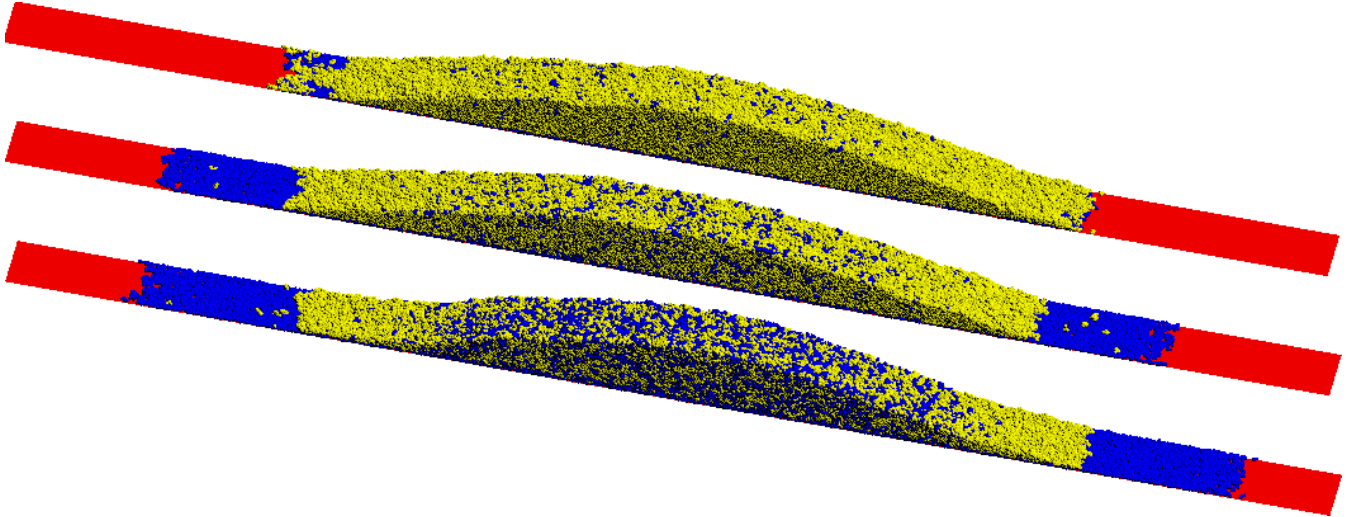}\end{center}

\caption{\label{cap:compositions} Profiles of three binary droplets of chain
length $N=10$ with $\varepsilon_{w}=2.0\,\varepsilon$ (yellow) and
$\varepsilon_{w}=3.0\,\varepsilon$ (blue). The mole fractions of
the strongly wetting component are $x_{wet}=0.10$ (top), $0.25$
(middle), and $0.50$ (bottom). $t=60\,000\,\tau$, $\gamma_{L}^{s}=10.0\,\tau^{-1}$.
The substrate (red) has a length of $800\,\sigma$ and a width of
$40\,\sigma$ in each profile.}
\end{figure}

\begin{figure}
\begin{center}\includegraphics[%
  clip,
  width=3.25in,
  keepaspectratio]{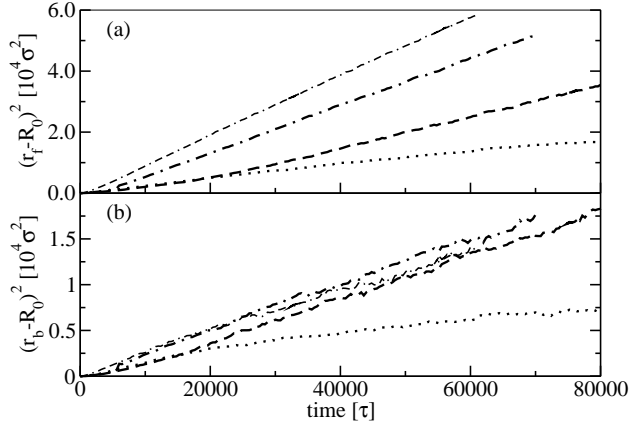}\end{center}

\caption{\label{cap:radcomp} Contact radii of the precursor foot (a) and
bulk (b) for the droplets composed of $x_{wet}=0$ (dotted curves),
$x_{wet}=0.10$ (dashed curves), $0.25$ (dot-dashed curves), and
$0.50$ (dot-dash-dashed curves). $\varepsilon_{w}=2.0\,\varepsilon$
for the weakly wetting component and $\varepsilon_{w}=3.0\,\varepsilon$
for the strongly wetting component. $N=10$, $\gamma_{L}^{s}=10.0\,\tau^{-1}$.}
\end{figure}

The ability to form a film from a non- or weakly wetting polymer by
adding a wetting component is of significant practical interest. To
analyze the effects of the concentration of the two components on
the spreading dynamics, we simulate droplets containing a strongly
wetting component ($\varepsilon_{w}=3.0\,\varepsilon$) at different
mole fractions, $x_{wet}=0.10$, $0.25$, and $0.50$ with a weakly
wetting component ($\varepsilon_{w}=2.0\,\varepsilon$). Since $N=10$
for both, the only difference in the two components is the surface
interaction. Profiles of the spreading droplets after $60\,000\,\tau$
are shown in Figure \ref{cap:compositions}. In each case, a monolayer
of the strongly wetting component forms on the surface and spreads
outward. The rest of the drop then spreads on this layer. However,
for $x_{wet}=0.10$, the surface monolayer never becomes fully homogeneous
as a segment of the weakly wetting component remains at the edge of
the foot and continues to get pushed outward as the droplet spreads.
The size of the depletion region at the droplet edge increases as
$x_{wet}$ decreases since less of the strongly wetting polymer is
available to form the precursor foot. The rate of spreading of the
precursor foot decreases as $x_{wet}$ decreases because the supply
rate of the strongly wetting material from the bulk to the surface
limits the spreading rate of the precursor foot. This is more clearly
demonstrated in Fig. \ref{cap:radcomp} where the contact radii of
both the foot and bulk regions are shown as a function of time for
each of the three compositions and compared to previous results\cite{HGW:PRE:04}
for $x_{wet}=0$. Figure \ref{cap:radcomp}a shows that the precursor
foot spreading rate decreases as $x_{wet}$ is reduced. The effective
diffusion constant of the precursor foot, $D_{eff}=\frac{1}{2}\frac{dr_{f}^{2}}{dt}$,
decreases from $4.7$ to $3.8$, $2.6$ and $0.6\,\sigma^{2}/\tau$
as $x_{wet}$ decreases from $0.50$ to $0.25$, $0.10$ and $0.0$,
respectively. Figure \ref{cap:radcomp}b shows no significant effect
of $x_{wet}$ on the bulk droplet spreading rate for the duration
of the simulation for $x_{wet}=0.5$ to $x_{wet}=0.1$. However, there
is a strong effect on the spreading rate when increasing the concentration
from $x_{wet}=0.0$ to $x_{wet}=0.1$. In all of these cases, the
bulk contact radius spreads as $t^{x}$ with $x\cong1/5$,\cite{HGW:PRE:04}
which is expected from the kinetic model for spreading in the cylindrical
geometry used here.

\begin{figure}
\begin{center}\includegraphics[%
  clip,
  width=3.25in
  ]{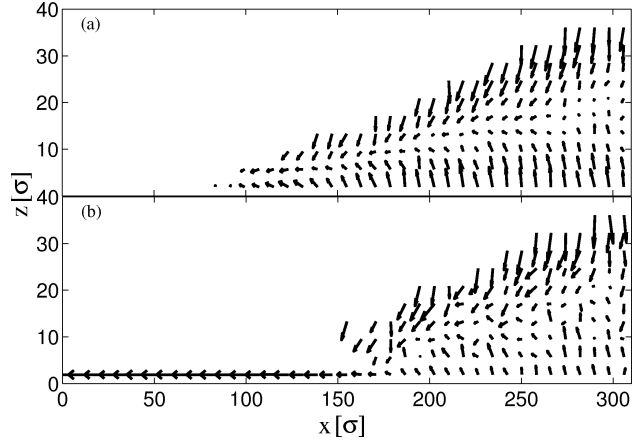}\end{center}

\caption{\label{cap:vel6} Velocity profiles of half of the droplet showing
the weakly wetting component, $\varepsilon_{w}=2.0\,\varepsilon$
(a), and the strongly wetting component, $\varepsilon_{w}=3.0\,\varepsilon$
(b), for the $x_{wet}=0.25$ binary droplet of chain length $N=10$
with $\gamma_{L}^{s}=10.0\,\tau^{-1}$ at $t=70\,000\,\tau$.}
\end{figure}

The velocity profiles of the strongly wetting and weakly wetting components
at $t=70\,000\,\tau$ are shown in Fig. \ref{cap:vel6} for $x_{wet}=0.25$.
After the foot has extended and a depletion region has formed at the
edge of the droplet, Fig. \ref{cap:vel6}a shows that the weakly wetting
material moves down from the liquid/vapor surface and up from the
solid/liquid surface to allow the monolayer of the strongly wetting
material to form. Very little motion is seen in the center of the
droplet. In Fig. \ref{cap:vel6}b, the strongly wetting material shows
similar behavior at the liquid/vapor surface, but the data is noisier
due to the smaller mole fraction of strongly wetting material. This
material moves through the depletion region at the edge of the droplet
and onto the precursor foot where it rapidly diffuses outward. The
same qualitative behavior is observed for different values of $x_{wet}$,
though the data becomes increasingly noisy as $x_{wet}$ decreases.

\section{Conclusions\label{sec:conclusions}}

The spreading dynamics of binary polymer nanodroplets are studied
using molecular dynamics simulation. We demonstrate qualitative differences
in the spreading behavior of binary droplets due to differences in
surface interaction strength, polymer chain length, and composition.

When the two droplet components differ in surface interaction strength,
the more strongly wetting component forms a monolayer film on the
surface even when both materials are either above or below the wetting
transition. For cases that also produce a rapidly spreading precursor
foot, a depletion region of the more strongly wetting component forms
starting at the edge of the droplet since the interdiffusion rate
in the droplet is slower than the rate at which material is withdrawn
into the precursor foot.

For differences in the polymer chain length, the monolayer film beneath
the droplet is composed of an equal amount of short chain and long
chain monomers. This is true even when one component (the shorter
chain length) is above the wetting transition and the other is not.
In this case, the precursor foot is composed primarily of the wetting
component. When both components are above the wetting transition,
the surface monolayer and precursor foot are both composed of an equal
amount of short and long chain monomers.

The formation of a monolayer of the more strongly wetting component
is studied in greater detail by considering different compositions
of the mixture of strongly wetting and weakly wetting components.
For each case, a monolayer of the strongly wetting component forms
on the surface and spreads outward. The spreading rate of this precursor
foot decreases as the concentration of the strongly wetting component
is reduced, but the spreading rate of the bulk droplet shows no significant
dependence on the droplet composition until all of the more strongly
wetting component is removed. Velocity profiles show that the material
for the precursor foot is supplied by the wetting material near the
liquid/vapor surface as it diffuses through the depletion region and
onto the substrate.

\section*{Acknowledgements}

Sandia is a multiprogram laboratory operated by Sandia Corporation,
a Lockheed Martin Company, for the United States Department of Energy's
National Nuclear Security Administration under Contract No. DE-AC04-94AL85000.

\end{document}